\documentclass[aps,pre,superscriptaddress,twocolumn]{revtex4}
\usepackage{amssymb}
\usepackage{amsfonts}
\usepackage{amsmath}
\usepackage{graphicx}

\begin{document}

\title{First passage time for superstatistical Fokker-Planck models}
\author{Adri\'an A. Budini}
\affiliation{Consejo Nacional de Investigaciones Cient\'{\i}ficas y T\'{e}cnicas
(CONICET), Centro At\'{o}mico Bariloche, Avenida E. Bustillo Km 9.5, (8400)
Bariloche, Argentina, and Universidad Tecnol\'{o}gica Nacional (UTN-FRBA),
Fanny Newbery 111, (8400) Bariloche, Argentina}
\author{Manuel O. C\'aceres}
\affiliation{Centro At\'{o}mico Bariloche, CNEA, Instituto Balseiro and CONICET, 8400
Bariloche, Argentina.}
\date{\today }

\begin{abstract}
The first passage time (FPT) problem is studied for superstatistical models
assuming that the mesoscopic system dynamics is described by a Fokker-Planck
equation. We show that all moments of the random intensive parameter
associated to the superstatistical approach can be put in one-to-one
correspondence with the moments of the FPT. For systems subjected to an
additional uncorrelated external force, the same statistical information is
obtained from the dependence of the FPT-moments on the external force. These
results provide an alternative technique for checking the validity of
superstatistical models. As an example, we characterize the mean FPT for a
forced Brownian particle.
\end{abstract}

\maketitle

%\pacs{ 02.50.Ga, 03.67.Mn, 05.40.Fb}

\section{Introduction}

Superstatistics is an approximation that allows to modeling complex
nonequilibrium system dynamics \cite{beckcohen}. It applies when temporal or
inhomogeneous spatial fluctuations of the environment occur on a large time
scale, while the local system dynamics relaxes in a faster time scale \cite%
{tres}, staying in equilibrium for some time. Hence, its dynamics follows
from a double average (superstatistics) consisting in a local equilibrium
probability distribution which in turn depends on a \textit{random intensive
parameter} associated to the environment fluctuations. A central feature of
this approach is the possibility of modeling non-Gibbsian equilibrium-like
system distributions~\cite{touchette}.

At the beginning of its formulation, superstatistics was applied in the
context of turbulent Lagrangian dynamics \cite{lagrange,turbu,straeten}.
Nevertheless, since that, it has been also used for studying diverse kind of
complex systems such as ultracold gases \cite{rous}, quantum entanglement in
Ising-like models \cite{oura}, bacterial DNA architecture \cite{DNA},
migration of tumor cells \cite{mark}, plasma physics \cite{tribeche},
nanoscale electromechanical systems \cite{morales}, work fluctuation
theorems \cite{work}, and market signals \cite{log} just to name a few \cite%
{abul,general,trafic,relativistic,plastino,brownian,BeckApplication}.

Many efforts were also devoted to deriving the superstatistical approach
from underlying descriptions such as entropic principles \cite%
{vander,abe,tsallis,cristian}, and dynamical ones based on underlying
Fokker-Planck equations that include the intensive parameter as a stochastic
dynamical variable \cite{fokkerSuper}. More recently, multiscale and
hierarchical systems \cite{jerarquia,salazar} were also proposed, as well as
Langevin dynamics with diffusing diffusivity \cite{seno}. Theoretical
results of these kinds are of relevance because they allow to exploring
under which conditions superstatistical dynamics may apply. With a similar
motivation, the main goal of this paper is to introduce an extra criterion
for checking the validity of superstatistical models. We study the\textit{\
first passage time} (FPT) problem for dynamics where this approach applies.

The FPT measures when a system of interest crosses a given boundary or
threshold value for the first time. Its statistics has played an important
role in many disciplines \cite{Redner,Kampen,Gardiner,Caceres}. Its
properties has been scarcely studied in the context of the superstatistical
approach \cite{brownian}. We show that the statistics of the FPT can be put
in-one-to-one correspondence with the statistics of the random intensive
parameter of the superstatistical model. The same information can be
obtained by driving the system with an external uncorrelated force. Added to
their theoretical interest, these results provide an alternative technique
for measuring the underlying environment fluctuations and checking in
consequence the validity of a superstatistical approximation.

The paper is outlined as follows. In Sec. II we calculate the FPT statistics
for both unforced and forced Fokker-Planck dynamics. In Sec. III we apply
the main results to a superstatistical (forced) Brownian particle. Sec. IV
is devoted to the conclusions. Calculus details that support the main
results are found in the Appendixes.

\section{Superstatistical model and first passage time statistics}

The FPT problem is sets by a domain $\mathcal{D}$ and its frontier $\partial 
\mathcal{D},$ which defines the first passage condition. In an
\textquotedblleft experimental\ setup,\textquotedblright\ the domain $%
\mathcal{D}$ may be fixed in \textquotedblleft space\textquotedblright\ or
alternatively, it may be located following the \textquotedblleft system
position.\textquotedblright\ The system relaxes in a faster time scale.
Thus, inside the domain $\mathcal{D},$ and during the measurement of the
FPT, we can assume that the environment is characterized by a (random)
time-independent \textquotedblleft intensive parameter\textquotedblright\ $%
\beta ,$ which modifies the system dynamics. We consider a system of
arbitrary dimensionality, and assume that its mesoscopic local dynamics
(during the measurement time) is described by a probability distribution $%
P_{t}(\mathbf{y}|\mathbf{x})$ which obeys a Fokker-Planck equation%
\begin{equation}
\frac{\partial }{\partial t}P_{t}(\mathbf{y}|\mathbf{x})=\frac{1}{\beta }%
\mathbb{L}_{0}P_{t}(\mathbf{y}|\mathbf{x}).
\end{equation}%
Here, $\mathbb{L}_{0}$ is the Fokker-Planck differential operator, where $%
\mathbf{y}$ labels the relevant system\ coordinate, while $\mathbf{x}\in 
\mathcal{D}$ is its value at the initial time $t_{0}=0.$

The previous description is consistent if each measurement of the FPT is
performed at separate time intervals larger than the correlation time of the
environment fluctuations. Thus, in agreement with the superstatistical
approach \cite{seno}, the value of $\beta $ in each measurement assumes a
different random value. It probability density $p(\beta )$ remains
unspecified.

As is well known, for stochastic processes that obey a Fokker-Planck
equation, the \textit{mean} FPT obey a differential equation, named as
Dynkin equation \cite{Dynkin}. It is defined by the adjoint Fokker-Planck
operator, where the spatial coordinate labels the initial system position 
\cite{Kampen,Gardiner,Caceres}. Hence, for each \textquotedblleft
realization\textquotedblright\ of $\beta ,$ the mean FPT $T^{(1)}(\mathbf{x}%
) $ obeys%
\begin{equation}
\beta ^{-1}\mathbb{L}_{0}^{\dagger }T^{(1)}(\mathbf{x})=-1,
\label{Dynkin_Mean}
\end{equation}%
where $\mathbb{L}_{0}^{\dagger }$ is the adjoint Fokker-Planck differential
operator. This equation must to be solved for $\mathbf{x\in }$ $\mathcal{D},$
where the domain $\mathcal{D}$ and its frontier $\partial \mathcal{D}$
define the first passage condition. Hence, $T^{(1)}(\partial \mathcal{D})=0.$
By denoting the average over $p(\beta )$ with $\langle \cdots \rangle ,$ the
previous equation straightforwardly leads to%
\begin{equation}
\mathbb{L}_{0}^{\dagger }\langle T^{(1)}(\mathbf{x})\rangle =-\langle \beta
\rangle .  \label{FirstPromedio}
\end{equation}%
Hence, the mean FPT is the same than in the absence of $\beta $%
-fluctuations, but with an effective value $\beta \rightarrow \langle \beta
\rangle .$ Now, we explore if a similar relation is valid for higher moments
of the FPT.

In general, higher moments $T^{(n)}(\mathbf{x})$ obey the recurrence
relation \cite{Gardiner,Caceres}%
\begin{equation}
\beta ^{-1}\mathbb{L}_{0}^{\dagger }T^{(n)}(\mathbf{x})=-nT^{(n-1)}(\mathbf{x%
}).  \label{recurrence}
\end{equation}%
Given that $T^{(n)}(\mathbf{x})$ and $\beta $ are correlated, in this case
averaging with $p(\beta )$ is not a trivial task. Techniques arising in
disordered systems \cite{garcia,pury} may be proposed for performing the
average. Nevertheless, a simpler solution is available. First, we notice
that Eq. (\ref{recurrence}) can formally be solved as $(\mathbb{L}%
_{0}^{\dagger })^{n}T^{(n)}(\mathbf{x})=(-1)^{n}n!\beta ^{n},$ where the
boundary conditions of this differential equation are $(\mathbb{L}%
_{0}^{\dagger })^{m}T^{(n)}(\partial \mathcal{D})=0,$ $\forall m\leq n.$
This solution says us that $T^{(n)}(\mathbf{x})$ only depends on $\beta
^{n}. $ Hence, averaging over $\beta $ becomes an easy task,%
\begin{equation}
(\mathbb{L}_{0}^{\dagger })^{n}\langle T^{(n)}(\mathbf{x})\rangle
=(-1)^{n}n!\langle \beta ^{n}\rangle .  \label{solution}
\end{equation}%
This result generalizes Eq. (\ref{FirstPromedio}). It gives us one of the
central results of this paper: as the $n$-moment $\langle T^{(n)}(\mathbf{x}%
)\rangle $\ of the FPT \textit{only} depends on the $n$-moment of $\langle
\beta ^{n}\rangle ,$ their measurement allows to determine the full
statistic (moments) of the random intensive parameter $\beta .$ The
dependence on $\mathbf{x}$ only takes into account the geometry of the
problem. Notice that the previous equation can be rewritten as%
\begin{equation}
\langle T^{(n)}(\mathbf{x})\rangle =T^{(n)}(\mathbf{x})|_{\beta
^{n}\rightarrow \langle \beta ^{n}\rangle }.
\end{equation}%
Hence, any moment $\langle T^{(n)}(\mathbf{x})\rangle $ can be obtained from
the standard solution $T^{(n)}(\mathbf{x})$ under the replacement $\beta
^{n}\rightarrow \langle \beta ^{n}\rangle .$ In contrast with Eq. (\ref%
{solution}), this result, jointly with Eq. (\ref{recurrence}), provides a
simpler technique for calculating $\langle T^{(n)}(\mathbf{x})\rangle .$

\subsection*{External forcing}

Now, we consider that an external driving force is applied over the system.
This situation is of interest from an experimental point of view. In fact,
the external force may delay the exit time, or even it may \textquotedblleft
localize\textquotedblright\ the system in the surroundings of the domain~$%
\mathcal{D}.$

We assume that the external force has not any correlation with the
environment fluctuations. Hence, the Fokker-Planck equation becomes%
\begin{equation}
\frac{\partial }{\partial t}P_{t}(\mathbf{y}|\mathbf{x})=(\varepsilon 
\mathbb{L}_{f}+\beta ^{-1}\mathbb{L}_{0})P_{t}(\mathbf{y}|\mathbf{x}).
\end{equation}%
Here, $\mathbb{L}_{f}$ is the contribution induced by the external force,
while the parameter $\varepsilon $ measures its strength. For each value of $%
\beta ,$ the FPT $n$-moments $T^{(n)}(\mathbf{x})$ obey the equation%
\begin{equation}
(\varepsilon \mathbb{L}_{f}^{\dagger }+\beta ^{-1}\mathbb{L}_{0}^{\dagger
})T^{(n)}(\mathbf{x})=-nT^{(n-1)}(\mathbf{x}).  \label{RecursivaForced}
\end{equation}%
In this case, averaging over $\beta $ is also a nontrivial task \cite{pury}.
Nevertheless, as in the previous case, Eq. (\ref{RecursivaForced}) can
formally be solved as%
\begin{equation}
(\varepsilon \mathbb{L}_{f}^{\dagger }+\beta ^{-1}\mathbb{L}_{0}^{\dagger
})^{n}T^{(n)}(\mathbf{x})=(-1)^{n}n!.  \label{TnDynkin}
\end{equation}%
For solving this equation we propose a series solution in the strength
force, $T^{(n)}(\mathbf{x})=\sum\nolimits_{j=0}^{\infty }\varepsilon
^{j}\beta ^{n+j}\tau _{j}(\mathbf{x}),$ where the set of functions $\{\tau
_{j}(\mathbf{x})\}_{j=0}^{\infty }$ (\textit{defined for each }$n$) satisfy
the same boundary conditions than $T^{(n)}(\mathbf{x}).$ This series
expansion implies that the set of functions $\{\tau _{j}(\mathbf{x}%
)\}_{j=0}^{\infty }$ do not depend on $\beta $ (Appendix A).\ Thus, we
arrive to the second main result,%
\begin{equation}
\langle T^{(n)}(\mathbf{x})\rangle =\sum\nolimits_{j=0}^{\infty }\varepsilon
^{j}\langle \beta ^{n+j}\rangle \tau _{j}(\mathbf{x}).  \label{Series}
\end{equation}%
This expression demonstrates that all moments $\langle \beta ^{n}\rangle $
can alternatively be determined by studying the dependence of $\langle
T^{(n)}(\mathbf{x})\rangle $ on the force strength $\varepsilon .$ On the
other hand, notice that the functions $\{\tau _{j}(\mathbf{x}%
)\}_{j=0}^{\infty }$ are the same that arise for a deterministic value of $%
\beta .$ They only depend on the geometry and symmetries of the problem.

From Eq. (\ref{Series}), for the mean FPT we get%
\begin{equation}
\langle T^{(1)}\rangle (\mathbf{x})\!=\!\langle \beta \rangle \tau _{0}(%
\mathbf{x})+\varepsilon \langle \beta ^{2}\rangle \tau _{1}(\mathbf{x}%
)+\varepsilon ^{2}\langle \beta ^{3}\rangle \tau _{2}(\mathbf{x})+\cdots ,
\label{FirstSeries}
\end{equation}%
where the functions $\{\tau _{j}(\mathbf{x})\}_{j=0}^{\infty }$ satisfy
(Appendix A)%
\begin{equation}
\mathbb{L}_{0}^{\dagger }\tau _{0}(\mathbf{x})=-1,\ \ \ \ \ \ \ \ \mathbb{L}%
_{0}^{\dagger }\tau _{j}(\mathbf{x})=-\mathbb{L}_{f}^{\dagger }\tau _{j-1}(%
\mathbf{x}),  \label{SeriesEquationMFPT}
\end{equation}%
which in fact do not depend on $\beta .$

\section{Symmetrically-forced Brownian motion}

The previous results are valid for models defined with arbitrary
Fokker-Planck operators $\mathbb{L}_{0}$\ and $\mathbb{L}_{f}.$ In order to
exemplify how they apply, we consider a diffusive particle whose \textit{%
velocity} relaxes in a faster time scale when compared with the environment
fluctuations. Thus, in the last time scale, its \textit{position} performs a
Brownian motion \cite{Kampen,Gardiner,Caceres}. In addition, we consider
that an external uncorrelated \textit{constant} \textit{force}, $\mathbf{F}(%
\mathbf{x})=\varepsilon \mathbf{n,}$ is applied over it. The (unit) vector $%
\mathbf{n}$ gives its direction in space. In consequence, the probability
density $P_{t}(\mathbf{y}|\mathbf{x})$ for its position $\mathbf{y}$ obeys
the (Smoluchowski) Fokker-Planck equation%
\begin{equation}
\frac{\partial }{\partial t}P_{t}(\mathbf{y}|\mathbf{x})=(\varepsilon 
\mathbf{\nabla \cdot n}+\beta ^{-1}\mathbf{\nabla }^{2})P_{t}(\mathbf{y}|%
\mathbf{x}),  \label{temperatura}
\end{equation}%
where $\mathbb{L}_{f}=\mathbf{\nabla }$ and $\mathbb{L}_{0}=\mathbf{\nabla }%
^{2}$ are the gradient and Laplacian operators in dimension $d=1,2,3.$
Notice that for this dynamics, $\beta ^{-1}$ becomes the \textit{diffusion
coefficient}. Therefore, $\beta $ may be associated with inverse-temperature
fluctuations of the environment.

The Dynkin equation for the mean FPT becomes%
\begin{equation}
(\varepsilon \mathbf{n\cdot \nabla }+\beta ^{-1}\mathbf{\nabla }^{2})T^{(1)}(%
\mathbf{x})=-1.  \label{T1DinkinDifusivo}
\end{equation}%
For simplifying and showing the main features of the problem, we consider
symmetrical domains and forces that do not break their symmetry. Hence, in $%
d=2,3$ the domain $\mathcal{D}$ is the inner region of a circle and a sphere
of radius $R$ respectively. The force is radial in both cases, which defines
the direction $\mathbf{n.}$ In $d=1,$ for mimicking the higher dimensional
cases, a reflecting boundary is taken at the origin. In all cases, the only
relevant (radial) initial coordinate is denoted with $\rho .$ Hence, the
frontier $\partial \mathcal{D}$\ is taken into account through the absorbing
boundary condition $T^{(1)}(\rho =R)=0.$

The solution of Eq. (\ref{T1DinkinDifusivo}) for arbitrary $\varepsilon $
and $\beta $ can be found in an exact way (Appendix B). After performing a
series expansion in $\varepsilon $ and averaging over $\beta ,$ we get the
exact expression%
\begin{equation}
\langle T^{(1)}(\rho )\rangle =\sum_{n=1}^{\infty }\frac{(-1)^{n+1}}{r_{d}(n)%
}\varepsilon ^{n-1}\langle \beta ^{n}\rangle (R^{n+1}-\rho ^{n+1}),
\label{T1Diffusive_Series}
\end{equation}%
where the function $r_{d}(n),$ depending on the space dimension $(d=1,2,3),$
read $r_{1}(n)=(n+1)n!,$ $r_{2}(n)=(n+1)(n+1)!,$ and $r_{3}(n)=(n+1)(n+2)!/2$
respectively. We notice that Eq. (\ref{T1Diffusive_Series}) is consistent
with the general result (\ref{FirstSeries}). In the limit $\varepsilon \ll 1,
$ it follows%
\begin{equation}
\langle T^{(1)}(\rho )\rangle \simeq \frac{\langle \beta \rangle }{a}%
(R^{2}-\rho ^{2})-\varepsilon \frac{\langle \beta ^{2}\rangle }{b}%
(R^{3}-\rho ^{3}),  \label{FirstOrder}
\end{equation}%
where the pairs of constants $(a,b),$ depending on the space dimension $%
(d=1,2,3),$ read $(2,6),$ $(4,18),$ and $(6,36)$ respectively.

In Fig. 1 we plot the mean FPT\ $\langle T^{(1)}(\rho )\rangle $ [Eq. (\ref%
{T1Diffusive_Series})] for a $3D$ Brownian particle, in the case in which
the $\beta -$fluctuations\ are Gamma distributed \cite{tres},%
\begin{equation}
p(\beta )=\frac{1}{\Gamma (\alpha )}\left( \frac{\alpha }{\langle \beta
\rangle }\right) ^{\alpha }\ \beta ^{\alpha -1}\exp \left( \frac{-\alpha
\beta }{\langle \beta \rangle }\right) ,  \label{gamma}
\end{equation}%
with $\alpha =3/2.$ The moments are%
\begin{equation}
\langle \beta ^{n}\rangle =2^{n}\frac{\Gamma (\alpha +n)}{\Gamma (\alpha )}%
\left( \frac{\langle \beta \rangle }{2\alpha }\right) ^{n}.
\end{equation}%
In addition, we plot the solution in absence of fluctuations $p(\beta
)=\delta (\beta -\langle \beta \rangle ),$ which is denoted as $T^{(1)}(\rho
).$ Time is measured in units of $T_{0}=\langle \beta \rangle R^{2}/6,$
which corresponds to the mean FPT starting at the origin in the undriven $3D$
case, Eq. (\ref{FirstOrder}) with $\rho =0$ and $\varepsilon =0$
respectively.%figura1%figura%figura%figura%figurav%figura%figura%figura%figura%figura%figura%figura%figura%figura%figurav%figura%figura%figura%figura%figura
%figura%figura%figura%figura%figurav%figura%figura%figura%figura%figura%figura%figura%figura%figura%figurav%figura%figura%figura%figura%figura
\begin{figure}[tbp]
\includegraphics[bb=47 870 717 1147,angle=0,width=8.7cm]{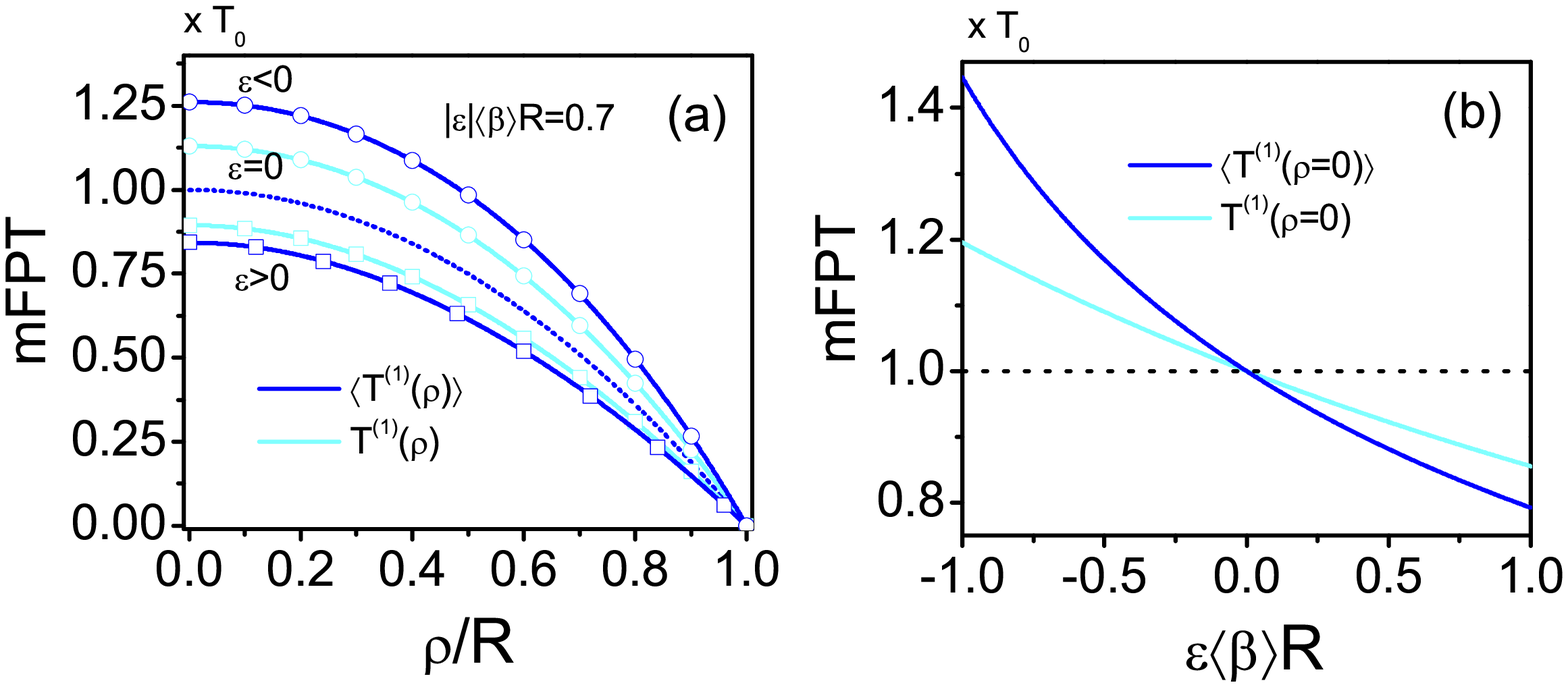}
\caption{Mean first passage time (mFPT) $\langle T^{(1)}(\protect\rho %
)\rangle $ in presence of a constant uncorrelated radial force in 3D [Eq. (%
\protect\ref{T1Diffusive_Series})] and Gamma distributed $\protect\beta -$%
fluctuations [Eq. (\protect\ref{gamma})]. $T^{(1)}(\protect\rho )$ is the
solution in absence of fluctuations, $p(\protect\beta )=\protect\delta (%
\protect\beta -\langle \protect\beta \rangle ).$ Time is measured in units $%
T_{0}=\langle \protect\beta \rangle R^{2}/6.$ (a) Dependence on the initial
position $\protect\rho ,$ for negative (circles), null (dotted line), and\
positive (squares) forces. The parameters fulfill $|\protect\varepsilon %
|\langle \protect\beta \rangle R=0.7.$ (b) Dependence with the external
dimensionless force $\protect\varepsilon \langle \protect\beta \rangle R$ at
the initial position $\protect\rho =0.$}
\end{figure}
%figura%figura%figura%figura%figurav%figura%figura%figura%figura%figura%figura%figura%figura%figura%figurav%figura%figura%figura%figura%figura
%figura%figura%figura%figura%figurav%figura%figura%figura%figura%figura%figura%figura%figura%figura%figurav%figura%figura%figura%figura%figura

In Fig. 1(a) we observe that, for any particular initial condition $\rho ,$
with respect to the undriven case (dotted line) a decreasing (increasing) of
the mean FPT is observed for outward $(\varepsilon >0,$ squares) (inwards, $%
\varepsilon >0,$ circles) radial forces. These two behaviors are expected
ones. Nevertheless, we notice that a crossover between\ $\langle
T^{(1)}(\rho )\rangle $ and $T^{(1)}(\rho )$ occurs when the force change of
sign. In fact $\langle T^{(1)}(\rho )\rangle \gtrless T^{(1)}(\rho )$ for $%
\varepsilon \lessgtr 0.$ This effect is independent of the specific value of 
$\varepsilon ,$ property confirmed in Fig. 1(b). In this case, we plot the
same objects as a function of the dimensionless force $\varepsilon \langle
\beta \rangle R$ at the initial condition $\rho =0.$

The crossover induced by the environment fluctuations (Fig. 1) can be
quantified from Eq. (\ref{FirstOrder}), which gives%
\begin{equation}
\langle T^{(1)}(\rho )\rangle -T^{(1)}(\rho )\simeq -\varepsilon \lbrack
\langle \beta ^{2}\rangle -\langle \beta \rangle ^{2}]\tau _{1}(\rho ),
\label{resta}
\end{equation}%
where $\tau _{1}(\rho )=(R^{3}-\rho ^{3})/36.$ Thus, the crossover around
small forces is governed by the cumulant $\langle \beta ^{2}\rangle -\langle
\beta \rangle ^{2}.$ This result gives us a procedure for checking the
environment fluctuations. From the undriven dynamics it is possible to
determine $\langle \beta \rangle .$ By subjecting the system to a small
constant force the mean FPT allows to measure the environment quadratic
fluctuations. In fact, in the forced case, $\langle T^{(1)}(\rho )\rangle $
cannot be fit with the single parameter $\langle \beta \rangle .$ From Eq. (%
\ref{T1Diffusive_Series}) we deduce that higher $n$-orders in $\varepsilon $
are weighted by $[\langle \beta ^{n}\rangle -\langle \beta \rangle ^{n}].$
Thus, the crossover gives us direct information about all centered moments
of the bath fluctuations.

We remark that the previous results do not depend on the specific studied
model. In fact, Eq. (\ref{resta}) remains valid in general. The only change
is the term $\tau _{1}(\rho )$ which takes into account the force and
geometry of the problem. For example, the same behaviors such as those shown
in Fig. 1 also arise for radial harmonic forces $\mathbf{F}(\mathbf{x}%
)=\varepsilon \mathbf{x}$ and Coulomb-like ones $\mathbf{F}(\mathbf{x}%
)=\varepsilon f(\mathbf{|\mathbf{x}}|)\mathbf{x/|\mathbf{x}}|,$ where $f(%
\mathbf{|\mathbf{x}}|)=1/\mathbf{|\mathbf{x}}|^{d-1},$ and $d=2,3$ is the
space dimensionality (Appendix B).

Forces that depend on position, as the previous ones, may be implemented at
a fixed position in space. Nevertheless, one may also be interested in
tracking the particle position while it diffuses in different spatial
regions of the environment. Thus, the successive domains $\mathcal{D}$
should be chosen around the (time-dependent) particle position. In this
case, a homogeneous force along space may be the more appropriate one, while 
$\mathcal{D}$ may be chosen as a \textquotedblleft square\textquotedblright\
surrounding the particle in each FPT measurement. This case can be
analytically characterized, and also confirms the proposed approach
(Appendix C). In fact, the previous results and conclusions remain valid
independently of the nature of the experimental setup chosen for measuring
the FPT.

We based our analysis on Eq. (\ref{temperatura}), which applies when the
environment induces \textit{diffusion coefficient} (temperature)
fluctuations. Alternatively, the environment may induce fluctuations in the 
\textit{dissipative coefficient} of the velocity dynamics. For the particle
position it becomes a global multiplicative constant \cite%
{Kampen,Gardiner,Caceres}. After a redefinition of the parameters, its
description can be recovered from Eq. (\ref{temperatura}) under the
replacement $\varepsilon \rightarrow \varepsilon \beta ^{-1}.$ In
consequence, the Dynkin equation (\ref{T1DinkinDifusivo}) assumes the
structure given by Eq. (\ref{Dynkin_Mean}). The mean FPT can therefore be
written as%
\begin{equation}
\langle T^{(1)}(\rho )\rangle =\langle \beta \rangle \sum_{n=1}^{\infty }%
\frac{(-1)^{n+1}}{r_{d}(n)}\varepsilon ^{n-1}(R^{n+1}-\rho ^{n+1}).
\end{equation}%
Comparing this expression with Eq. (\ref{T1Diffusive_Series}), we realize
that the measurement of the mean FPT in the forced case allows, in addition,
to discriminate between both situations, that is,\textit{\ diffusion }versus 
\textit{dissipative }coefficient fluctuations\textit{.} This property is
also valid for arbitrary external forces.

\section{Summary and Conclusions}

We have characterized the FPT problem for superstatistical models where the
mesoscopic local system dynamics is described by a Fokker-Planck operator.
In the first case, the (slow) environmental fluctuations are taken into
account by an intensive parameter that affects the total system dynamics. It
was shown that all moments of the random intensive parameter can be put in
one-to-one correspondence with the moments of the FPT. In addition, in the
second case, it was shown that the same statistical information can be
obtained by subjecting the system to an external force that is independent
of the environment fluctuations. The approach was exemplified with the
paradigmatic case of a superstatistical (forced) Brownian particle. In the
independent force case, depending on its action, a crossover between the
mean FPT in presence and absence of environmental fluctuations is observed.
Furthermore, we showed that external driving forces allows to discriminate
between\textit{\ }diffusion and dissipative coefficient fluctuations. These
properties are a fingerprint of the environment fluctuations which are found
whenever a superstatistical approach applies.

The obtained results provide two complementary experimental methods for
checking the consistence of the hypothesis (spatiotemporal scales) under
which a superstatistical modeling may applies. In fact, the superstatistical
approach leads to a precise structure of the FPT problem that goes beyond
the possible consistence with stationary non-Gibbsian equilibrium-like
distributions. Given the present technological advances in single particle
tracking, as well as the possibility of processing measurement signals of
diverse complex systems, the developed results provide a theoretical tool
that may be of relevance in different realistic experimental setups.

\section*{Acknowledgments}

A. A. B. acknowledge support from CONICET, Argentina. M.O.C. gratefully
acknowledge support received from CONICET, grant PIP 112-261501-00216 CO,
Argentina.

\appendix

\section{Solution for the first passage n-moment}

Here, we demonstrate that the $n$-moment of the first passage time (FPT) $%
T^{(n)}(\mathbf{x}),$ in the forced case, can be expressed as the series
expansion%
\begin{equation}
T^{(n)}(\mathbf{x})=\beta ^{n}\tau _{0}(\mathbf{x})+\varepsilon \beta
^{n+1}\tau _{1}(\mathbf{x})+\varepsilon ^{2}\beta ^{n+2}\tau _{2}(\mathbf{x}%
)+\cdots ,  \label{Tene}
\end{equation}%
which lead to Eq. (\ref{Series}). $T^{(n)}(\mathbf{x})$ satisfies the
recursive relation 
\begin{equation}
(\varepsilon \mathbb{L}_{f}^{\dagger }+\beta ^{-1}\mathbb{L}_{0}^{\dagger
})T^{(n)}(\mathbf{x})=-nT^{(n-1)}(\mathbf{x}),
\end{equation}%
which is formally solved by%
\begin{equation}
(\varepsilon \mathbb{L}_{f}^{\dagger }+\beta ^{-1}\mathbb{L}_{0}^{\dagger
})^{n}T^{(n)}(\mathbf{x})=(-1)^{n}n!.  \label{DynkinRepetida}
\end{equation}%
The differential operator is rewritten as%
\begin{equation}
(\varepsilon \mathbb{L}_{f}^{\dagger }+\beta ^{-1}\mathbb{L}_{0}^{\dagger
})^{n}=\sum\nolimits_{i=0}^{n}\frac{\varepsilon ^{i}}{\beta ^{n-i}}\mathbb{S}%
_{i},
\end{equation}%
where $\{\mathbb{S}_{i}\}_{i=0}^{n}$ are operators that can be written in
terms of $\mathbb{L}_{0}^{\dagger }$ and $\mathbb{L}_{f}^{\dagger }.$
Applying the previous expression to Eq.~(\ref{Tene}), after solving each
order in the $\varepsilon ,$ Eq. (\ref{DynkinRepetida}) leads to a set of
consistent equations for the functions $\{\tau _{j}(\mathbf{x}%
)\}_{j=0}^{\infty }.$ We explicitly get%
\begin{equation}
\mathbb{S}_{0}[\tau _{0}(\mathbf{x})]=(-1)^{n}n!,
\end{equation}%
while for $1\leq i<n$%
\begin{equation}
\sum\nolimits_{j=0}^{i}\mathbb{S}_{j}[\tau _{i-j}(\mathbf{x})]=0,\ \ \ \ \
1\leq i<n.
\end{equation}%
For $i\geq n$ it follows%
\begin{equation}
\sum\nolimits_{j=0}^{n}\mathbb{S}_{j}[\tau _{i-j}(\mathbf{x})]=0,\ \ \ \ \
n\leq i.
\end{equation}%
These set of equations do not include the parameter $\beta .$ Therefore, the
series Eq. (\ref{Tene}) gives a consistent solution of Eq.~(\ref%
{DynkinRepetida}). In fact, after solving $\tau _{0}(\mathbf{x})$ one can
obtain $\tau _{1}(\mathbf{x}),$ and so on. The posterior average over $\beta 
$ is straightforward.

For $n=1,$ after taking $\mathbb{S}_{0}=\mathbb{L}_{0}^{\dagger }$ and $%
\mathbb{S}_{1}=\mathbb{L}_{f}^{\dagger },$ the previous relations lead to%
\begin{equation}
\mathbb{L}_{0}^{\dagger }\tau _{0}(\mathbf{x})=-1,\ \ \ \ \ \ \ \ \mathbb{L}%
_{0}^{\dagger }\tau _{j}(\mathbf{x})=-\mathbb{L}_{f}^{\dagger }\tau _{j-1}(%
\mathbf{x}).
\end{equation}

For $n=2,$ the equations are%
\begin{equation}
\mathbb{S}_{0}[\tau _{0}(\mathbf{x})]=2,
\end{equation}%
while for $\tau _{1}(\mathbf{x})$%
\begin{equation}
\mathbb{S}_{0}[\tau _{1}(\mathbf{x})]+\mathbb{S}_{1}[\tau _{0}(\mathbf{x}%
)]=0.
\end{equation}%
For $\tau _{j}(\mathbf{x})$ with $j\geq 2,$%
\begin{equation}
\mathbb{S}_{0}[\tau _{j}(\mathbf{x})]+\mathbb{S}_{1}[\tau _{j-1}(\mathbf{x}%
)]+\mathbb{S}_{2}[\tau _{j-2}(\mathbf{x})]=0.
\end{equation}%
The operators are%
\begin{equation}
\mathbb{S}_{0}=(\mathbb{L}_{0}^{\dagger })^{2},\ \ \ \ \mathbb{S}_{1}=%
\mathbb{L}_{0}^{\dagger }\mathbb{L}_{f}^{\dagger }+\mathbb{L}_{f}^{\dagger }%
\mathbb{L}_{0}^{\dagger },\ \ \ \ \mathbb{S}_{2}=(\mathbb{L}_{f}^{\dagger
})^{2}.
\end{equation}

\section{Exact solutions for the mean FPT for radial forces}

In presence of radial forces, $\mathbf{F}(\mathbf{x})=\varepsilon f(\mathbf{|%
\mathbf{x}}|)\mathbf{x/|\mathbf{x}}|,$ the Fokker-Planck equation is
(uncorrelated force)%
\begin{equation}
\frac{\partial }{\partial t}P_{t}(\mathbf{y}|\mathbf{x})=\left( \varepsilon 
\mathbf{\nabla \cdot }\frac{\mathbf{y}}{\mathbf{|y}|}f(\mathbf{|y}|)+\beta
^{-1}\mathbf{\nabla }^{2}\right) P_{t}(\mathbf{y}|\mathbf{x}).
\end{equation}%
Dynkin equation hence becomes%
\begin{equation}
\left( \varepsilon f(\mathbf{|\mathbf{x}}|)\frac{\mathbf{x}}{\mathbf{|%
\mathbf{x}}|}\cdot \mathbf{\nabla }+\beta ^{-1}\mathbf{\nabla }^{2}\right)
T^{(1)}(\mathbf{x})=-1.  \label{DynkinRadial}
\end{equation}%
This equation can be solved for different space dependences $f(\mathbf{|%
\mathbf{x}}|)$ and configurations of interest. In each case, the parameter $%
\varepsilon $ changes its units.

\subsection{Constant radial forces}

For constant radial forces, it follows $f(\mathbf{|\mathbf{x}}|)=1.$
Solutions of Dynkin equation depend on the space dimensionality and the
chosen domain $\mathcal{D}.$

\subsubsection{Mean FPT\ in $1D$}

In the unidimensional case the Brownian particle moves on a line. Its
position is denoted by $\rho .$ Eq. (\ref{DynkinRadial}) becomes%
\begin{equation}
\varepsilon \frac{\partial T^{(1)}(\rho )}{\partial \rho }+\beta ^{-1}\frac{%
\partial ^{2}T^{(1)}(\rho )}{\partial \rho ^{2}}=-1.  \label{1D}
\end{equation}%
We consider the passage problem in the domain $\mathcal{D}=[0,R].$ Hence, $%
T^{(1)}(R)=0.$ For simplicity, we consider that the origin is a reflecting
boundary $\partial T^{(1)}(\rho )/\partial \rho |_{\rho =0}=0.$ In addition,
this election mimics the boundary conditions in higher dimensions.

By a direct integration, and after imposing the boundary conditions, the
solution of (\ref{1D}) can be written as%
\begin{equation}
T^{(1)}(\rho )=\frac{1}{\varepsilon }(R-\rho )-\frac{1}{\varepsilon }%
\int_{\rho }^{R}\exp [-\varepsilon \beta \rho ^{\prime }]d\rho ^{\prime }.
\end{equation}%
By performing a series expansion of the integral contribution, terms
proportional to $(1/\varepsilon )$ cancels out, while the remaining ones
recover Eq. (\ref{T1Diffusive_Series}).

\subsubsection{Mean FPT\ in $2D$}

In the plane, the domain $\mathcal{D}$ is taken as a circle of radius $R.$
The external force is directed in the radial direction, whose coordinate is
denoted by $\rho .$ Given the symmetry of the problem, Eq. (\ref%
{DynkinRadial}) is%
\begin{equation}
\varepsilon \frac{\partial T^{(1)}(\rho )}{\partial \rho }+\frac{\beta ^{-1}%
}{\rho }\frac{\partial }{\partial \rho }\left( \rho \frac{\partial
T^{(1)}(\rho )}{\partial \rho }\right) =-1.  \label{D2_Cte}
\end{equation}%
This equation can be solved by finding two homogeneous solutions, $%
Y_{1}(\rho )$ and $Y_{2}(\rho ),$ and writing a particular one $Y_{p}(\rho )$
as a function of them \cite{morse}, given rise to $T^{(1)}(\rho
)=c_{1}Y_{1}(\rho )+c_{2}Y_{2}(\rho )+Y_{p}(\rho ).$ We get $Y_{1}(\rho )=1,$
$Y_{2}(\rho )=\int^{\rho }d\rho ^{\prime }\exp (-\varepsilon \beta \rho
^{\prime })/\rho ^{\prime },$ and $Y_{p}(\rho )=-\varepsilon ^{-1}[\rho
-(\varepsilon \beta )^{-1}\ln (\rho )].$ The two indeterminate constants, $%
c_{1}$ and $c_{2},$ are chosen for satisfying the boundary condition $%
T^{(1)}(R)=0$ and for avoiding a divergence at the origin. The solution then
reads%
\begin{equation}
T^{(1)}(\rho )=\frac{1}{\varepsilon }(R-\rho )-\frac{1}{\varepsilon
^{2}\beta }\int_{\rho }^{R}\frac{1-\exp [-\varepsilon \beta \rho ^{\prime }]%
}{\rho ^{\prime }}d\rho ^{\prime }.
\end{equation}%
By performing a series expansion of the integral contribution, Eq. (\ref%
{T1Diffusive_Series}) is recovered.

\subsubsection{Mean FPT\ in $3D$}

In the three dimensional case, the domain $\mathcal{D}$ is taken as a sphere
of radius $R.$ Given the symmetry of the problem, the solution only depends
on the radial coordinate $\rho .$ The Dynkin differential equation becomes
[Eq. (\ref{DynkinRadial})]%
\begin{equation}
\varepsilon \frac{\partial T^{(1)}(\rho )}{\partial \rho }+\frac{\beta ^{-1}%
}{\rho ^{2}}\frac{\partial }{\partial \rho }\left( \rho ^{2}\frac{\partial
T^{(1)}(\rho )}{\partial \rho }\right) =-1,  \label{3DCte}
\end{equation}%
with boundary condition $T^{(1)}(R)=0.$ The solution can be found as in the
two-dimensional case, $T^{(1)}(\rho )=c_{1}Y_{1}(\rho )+c_{2}Y_{2}(\rho
)+Y_{p}(\rho ),$ with $Y_{1}(\rho )=1,$ $Y_{2}(\rho )=\int^{\rho }d\rho
^{\prime }\exp (-\varepsilon \beta \rho ^{\prime })/(\rho ^{\prime })^{2},$
and $Y_{p}(\rho )=\varepsilon ^{-3}\beta ^{-2}[2\rho ^{-1}+2(\varepsilon
\beta )\ln (\rho )-(\varepsilon \beta )^{2}\rho ].$ The exact solution is%
\begin{equation}
T^{(1)}(\rho )=\frac{1}{\varepsilon }(R-\rho )-\frac{2\beta ^{2}}{%
\varepsilon ^{3}}\int_{\rho }^{R}\frac{\exp [-\varepsilon \beta \rho
^{\prime }]-(1-\varepsilon \beta \rho ^{\prime })}{\rho ^{\prime 2}}d\rho
^{\prime }.
\end{equation}%
By performing a series expansion of the integral contribution, Eq. (\ref%
{T1Diffusive_Series}) is recovered.

\subsection{Harmonic radial forces}

Here, we consider harmonic radial forces, $f(\mathbf{|\mathbf{x}}|)=\mathbf{|%
\mathbf{x}}|.$

\subsubsection{Mean FPT\ in $2D$}

In this case, the domain $\mathcal{D}$ is also taken as a circle of radius $%
R.$ Given the symmetry of the problem, Eq. (\ref{DynkinRadial}) is%
\begin{equation}
\varepsilon \rho \frac{\partial T^{(1)}(\rho )}{\partial \rho }+\frac{\beta
^{-1}}{\rho }\frac{\partial }{\partial \rho }\left( \rho \frac{\partial
T^{(1)}(\rho )}{\partial \rho }\right) =-1.  \label{Harmonico2D}
\end{equation}%
Notice that this expression follow from Eq. (\ref{D2_Cte}) under the
replacement $\varepsilon \rightarrow \varepsilon \rho .$ The boundary
condition is $T^{(1)}(R)=0.$

The homogeneous solutions of Eq. (\ref{Harmonico2D}) are $Y_{1}(\rho )=1,$
and $Y_{2}(\rho )=\int^{\rho }d\rho ^{\prime }(1/\rho ^{\prime })\exp
(-\varepsilon \beta \rho ^{\prime 2}/2).$ The particular one is $Y_{p}(\rho
)=-\ln (\rho )/\varepsilon .$ After imposing the boundary condition, we get
the exact solution%
\begin{equation}
T^{(1)}(\rho )=\frac{1}{\varepsilon }\int_{\rho }^{R}\frac{1-\exp
[-\varepsilon \beta \rho ^{\prime 2}/2]}{\rho ^{\prime }}d\rho ^{\prime }.
\end{equation}%
By performing a series expansion of the integral contribution, and after
performing the average over $\beta ,$ it follows the series%
\begin{equation}
\langle T^{(1)}(\rho )\rangle =\sum_{n=1}^{\infty }\frac{(-1)^{n+1}}{r_{2}(n)%
}\varepsilon ^{n-1}\langle \beta ^{n}\rangle (R^{2n}-\rho ^{2n}).
\end{equation}%
Here, $r_{2}(n)=2^{n+1}n!n.$ In the limit $\varepsilon \ll 1,$ we get%
\begin{equation}
\langle T^{(1)}(\rho )\rangle \simeq \frac{\langle \beta \rangle }{4}%
(R^{2}-\rho ^{2})-\varepsilon \frac{\langle \beta ^{2}\rangle }{32}%
(R^{4}-\rho ^{4}).
\end{equation}

\subsubsection{Mean FPT\ in $3D$}

The domain $\mathcal{D}$ is a sphere of radius $R.$ Given the symmetry of
the problem, Dynkin equation (\ref{DynkinRadial}) becomes%
\begin{equation}
\varepsilon \rho \frac{\partial T^{(1)}(\rho )}{\partial \rho }+\frac{\beta
^{-1}}{\rho ^{2}}\frac{\partial }{\partial \rho }\left( \rho ^{2}\frac{%
\partial T^{(1)}(\rho )}{\partial \rho }\right) =-1,  \label{Harmonico3D}
\end{equation}%
with boundary condition $T^{(1)}(R)=0.$ This equation can be read from Eq. (%
\ref{3DCte}) under the replacement $\varepsilon \rightarrow \varepsilon \rho
.$

The homogeneous solutions of Eq. (\ref{Harmonico3D}) are $Y_{1}(\rho )=1,$
and $Y_{2}(\rho )=\int^{\rho }d\rho ^{\prime }(1/\rho ^{\prime 2})\exp
(-\varepsilon \beta \rho ^{\prime 2}/2).$ The particular one is $Y_{p}(\rho
)=-\beta \int^{\rho }d\rho ^{\prime }[\digamma (\rho ^{\prime })/\rho
^{\prime 2}]\exp (-\varepsilon \beta \rho ^{\prime 2}/2),$ where the
function $\digamma (\rho )$ is defined by the relation%
\begin{equation}
\frac{d}{d\rho }\digamma (\rho )=\rho ^{2}\exp \left( \frac{\varepsilon
\beta \rho ^{2}}{2}\right) .  \label{efe}
\end{equation}%
The solution $Y_{2}(\rho )$ is divergent at the origin. Thus, the final
exact solution of Eq. (\ref{Harmonico3D}) can be written as%
\begin{equation}
T^{(1)}(\rho )=\beta \int_{\rho }^{R}\frac{\exp [-\varepsilon \beta \rho
^{\prime 2}/2]}{\rho ^{\prime 2}}\digamma (\rho ^{\prime })d\rho ^{\prime }.
\label{SolHarmonico3D}
\end{equation}%
For performing a series expansion in $\varepsilon ,$ first we note that $%
\digamma (\rho ),$ from Eq. (\ref{efe}), can be written as%
\begin{equation}
\digamma (\rho )=\sum_{n=0}^{\infty }\frac{\varepsilon ^{n}\beta ^{n}\rho
^{2n+3}}{2^{n}n!(2n+3)}.
\end{equation}%
By inserting this expression into Eq. (\ref{SolHarmonico3D}), developing in
series the exponential factor, and after performing the average over $\beta
, $ we get%
\begin{equation}
\langle T^{(1)}(\rho )\rangle =\sum_{n=1}^{\infty }\frac{(-1)^{n+1}}{r_{3}(n)%
}\varepsilon ^{n-1}\langle \beta ^{n}\rangle (R^{2n}-\rho ^{2n}).
\label{Solsol3}
\end{equation}%
Here, $r_{3}(n)=(2n)!(1+2n)/[2^{n-1}(n-1)!].$ For $\varepsilon \ll 1,$%
\begin{equation}
\langle T^{(1)}(\rho )\rangle \simeq \frac{\langle \beta \rangle }{6}%
(R^{2}-\rho ^{2})-\varepsilon \frac{\langle \beta ^{2}\rangle }{60}%
(R^{4}-\rho ^{4}).
\end{equation}

In deriving Eq. (\ref{Solsol3}) we have used the product series result: $%
\sum_{i=0}^{\infty }a_{i}x^{i}\sum_{j=0}^{\infty
}b_{j}x^{j}=\sum_{k=0}^{\infty }c_{k}x^{k},$ with coefficient $%
c_{k}=\sum_{s=0}^{k}a_{s}b_{k-s}.$ Furthermore, the addition $%
\sum_{s=0}^{n}(-1)^{s}[s!(2s+3)(n-s)]^{-1}=\sqrt{\pi }/[4\Gamma (n+\frac{5}{2%
})],$ jointly with $\Gamma (n+1)=n\Gamma (n)$ and $\Gamma (\frac{1}{2}+n)=%
\sqrt{\pi }(2n)!/(4^{n}n!),$ where $\Gamma (x)$ is the Gamma function \cite%
{atlas}.

\subsection{Coulomb-like forces}

Here, we consider radial forces whose strength change in space as Coulomb
electrical forces do.

\subsubsection{Mean FPT\ in $2D$}

As an underlying physical model, we consider an electric charge uniformly
distributed on a (effectively infinite) cylinder of radius $R_{a}.$ The
charge generates an electrical field (force) which, over the plane
perpendicular to the cylinder, is directed in the radial direction. Its
amplitude decay as $f(\rho )=1/\rho .$ It is assumed that diffusion in the
direction parallel to the cylinder axe can be disregarded. Thus, the problem
is a bidimensional one. The domain $\mathcal{D}$ is taken as a circle of
radius $R_{b}>R_{a}.$ Given the symmetry of the problem, Eq. (\ref%
{DynkinRadial}) becomes%
\begin{equation}
\frac{\varepsilon }{\rho }\frac{\partial T^{(1)}(\rho )}{\partial \rho }+%
\frac{\beta ^{-1}}{\rho }\frac{\partial }{\partial \rho }\left( \rho \frac{%
\partial T^{(1)}(\rho )}{\partial \rho }\right) =-1.  \label{Coulomb2D}
\end{equation}%
This expression follows from Eq. (\ref{D2_Cte}) under the replacement $%
\varepsilon \rightarrow \varepsilon /\rho .$ We consider that the Brownian
particle cannot diffuse inside the cylinder. Hence, $(\partial /\partial
\rho )T^{(1)}(\rho )|_{\rho =R_{a}}=0.$ Furthermore, $T^{(1)}(R_{b})=0.$

The homogeneous solutions of Eq. (\ref{Coulomb2D}) are $Y_{1}(\rho )=1,$ and 
$Y_{2}(\rho )=\rho ^{-\varepsilon \beta },$ while the particular one reads $%
Y_{p}(\rho )=-\rho ^{2}/[2(2\beta ^{-1}+\varepsilon )].$ After imposing the
boundary conditions it follows $(R_{a}\leq \rho \leq R_{b}),$%
\begin{equation}
T^{(1)}(\rho )=\frac{(R_{b}^{2}-\rho ^{2})}{2(2\beta ^{-1}+\varepsilon )}+%
\frac{\beta ^{-1}R_{a}^{2+\varepsilon \beta }}{\varepsilon (2\beta
^{-1}+\varepsilon )}(R_{b}^{-\varepsilon \beta }-\rho ^{-\varepsilon \beta
}).  \label{C2SolAB}
\end{equation}%
In the limit $R_{a}\rightarrow 0,$ it reduces to $(R_{b}\rightarrow R)$%
\begin{equation}
T^{(1)}(\rho )=\frac{(R^{2}-\rho ^{2})}{2(2\beta ^{-1}+\varepsilon )}.
\label{C2SolUnRadio}
\end{equation}%
This last solution is only valid for $\varepsilon >-\beta /2,$ while for $%
\varepsilon \leq -\beta /2$ a divergence is obtained, $T^{(1)}(\rho )=\infty
.$ This singular property follows because for $\varepsilon \leq -\beta /2$
the origin lead to a non-regular boundary condition \cite{Kampen,Caceres}.

By performing a series expansion in $\varepsilon ,$ both Eq. (\ref{C2SolAB})
and (\ref{C2SolUnRadio}) confirm the series expansion found in the
manuscript. For the former one, we get%
\begin{eqnarray}
\langle T^{(1)}(\rho )\rangle &\simeq &\left[ \frac{R_{b}^{2}-\rho ^{2}}{4}+%
\frac{R_{a}^{2}\ln (\frac{\rho }{R_{b}})}{2}\right] \langle \beta \rangle
-\varepsilon \langle \beta ^{2}\rangle \\
&&\times \left\{ \frac{R_{b}^{2}-\rho ^{2}}{8}+\frac{R_{a}^{2}\ln (\frac{%
\rho }{R_{b}})[1+\ln (\frac{R_{b}\rho }{R_{a}^{2}})]}{4}\right\} ,  \notag
\end{eqnarray}%
while for the last one it follows%
\begin{equation}
\langle T^{(1)}(\rho )\rangle \simeq \frac{(R^{2}-\rho ^{2})}{4}\left(
\langle \beta \rangle -\varepsilon \frac{\langle \beta ^{2}\rangle }{2}%
+\varepsilon ^{2}\frac{\langle \beta ^{3}\rangle }{4}\cdots \right) .
\end{equation}

\subsubsection{Mean FPT\ in $3D$}

In this example, the electric charge is uniformly distributed on a sphere of
radius $R_{a}.$ This distribution generates a radial force with intensity $%
f(\rho )=1/\rho ^{2}.$ The domain $\mathcal{D}$ is taken as a sphere of
radius $R_{b}>R_{a}.$ Given the symmetry of the problem, Eq. (\ref%
{DynkinRadial}) becomes%
\begin{equation}
\frac{\varepsilon }{\rho ^{2}}\frac{\partial T^{(1)}(\rho )}{\partial \rho }+%
\frac{\beta ^{-1}}{\rho ^{2}}\frac{\partial }{\partial \rho }\left( \rho ^{2}%
\frac{\partial T^{(1)}(\rho )}{\partial \rho }\right) =-1,  \label{Coulomb3D}
\end{equation}%
which can be read from Eq. (\ref{3DCte}) under the replacement $\varepsilon
\rightarrow \varepsilon /\rho ^{2}.$ Given that the particle cannot diffuse
inside the sphere, it follows $(\partial /\partial \rho )T^{(1)}(\rho
)|_{\rho =R_{a}}=0,$ while $T^{(1)}(R_{b})=0.$

The solution of Eq. (\ref{Coulomb3D}) can be written as%
\begin{eqnarray}
T^{(1)}(\rho ) &=&c_{1}+c_{2}\exp (\varepsilon \beta /\rho )+\frac{%
(\varepsilon \rho -\beta ^{-1}\rho ^{2})}{6\beta ^{-2}}  \notag \\
&&+\frac{\varepsilon ^{2}\exp (\varepsilon \beta /\rho )}{6\beta ^{-3}}%
E_{i}(-\beta /\rho ),  \label{TcUN}
\end{eqnarray}%
where $E_{i}(z)\equiv -\int_{-z}^{\infty }dt\exp [-t]/t$ is the exponential
integral function. The undetermined constants $c_{1}$\ and $c_{2}$\ can be
obtained after imposing the boundary conditions. We notice that in the case $%
R_{a}=0,$ for avoiding a divergence at the origin, the constant $c_{2}$ must
to vanishes. In this limit case, only solutions for $\varepsilon >0$ are
admissible, while $T^{(1)}(\rho )=\infty $ for $\varepsilon <0.$

A series expansion in the parameter $\varepsilon $ of Eq. (\ref{TcUN}) leads%
\begin{eqnarray}
\langle T^{(1)}(\rho )\rangle \! &\simeq &\!\left[ \frac{R_{b}^{2}-\rho ^{2}%
}{6}-\frac{R_{a}^{3}(R_{b}-\rho )}{3R_{b}\rho }\right] \langle \beta \rangle
-\varepsilon \langle \beta ^{2}\rangle \\
&&\!\!\!\!\!\!\!\!\!\!\times \left\{ \frac{(R_{b}-\rho )[R_{b}^{2}\rho
^{2}+R_{a}^{3}(R_{b}+\rho )-3R_{a}^{2}R_{b}\rho ]}{6R_{b}^{2}\rho ^{2}}%
\right\} .  \notag
\end{eqnarray}%
For $R_{a}=0,$ $R_{b}\rightarrow R,$ the previous result reduces to%
\begin{equation}
\langle T^{(1)}(\rho )\rangle \simeq \langle \beta \rangle \frac{(R^{2}-\rho
^{2})}{6}-\varepsilon \frac{\langle \beta ^{2}\rangle }{6}(R-\rho ),
\end{equation}%
expression valid for $\varepsilon >0.$

\section{Exact solutions for the mean FPT for spatially homogeneous forces}

In this example the force is spatially homogeneous. We consider a $2D$
problem, where the coordinates are $(x,y).$ The domain $\mathcal{D}$ is
taken as a square of length $L,$ one of its vertices being located at the
origin of coordinates. The force is directed in the $x$-direction. Hence,
the Dynkin equation becomes%
\begin{equation}
\left[ \beta ^{-1}\left( \frac{\partial ^{2}}{\partial x^{2}}+\frac{\partial
^{2}}{\partial y^{2}}\right) +\varepsilon \frac{\partial }{\partial x}\right]
T^{(1)}(x,y)=-1,  \label{DynPlano}
\end{equation}%
with $T^{(1)}(0,y)=T^{(1)}(L,y)=T^{(1)}(x,0)=T^{(1)}(x,L)=0.$ This equation
can be solved by proposing the solution%
\begin{equation}
T^{(1)}(x,y)=\sum_{m=1}^{\infty }a_{m}\sin (k_{m}y)f_{m}(x),  \label{Sol}
\end{equation}%
where $k_{m}=m\pi /L,$ while the coefficients $\{a_{m}\}$ are defined by the
relation%
\begin{equation}
\sum_{m=1}^{\infty }a_{m}\sin (k_{m}y)=1,\ \ \ \ \ \ y\in (0,L),
\end{equation}%
leading to $a_{m}=4/(m\pi )$ for odd $m,$ while $a_{m}=0$ for even $m.$
Inserting Eq. (\ref{Sol}) into Dynkin equation (\ref{DynPlano}) it follows%
\begin{equation}
\left[ \beta ^{-1}\left( \frac{\partial ^{2}}{\partial x^{2}}%
-k_{m}^{2}\right) +\varepsilon \frac{\partial }{\partial x}\right]
f_{m}(x)=-1.
\end{equation}%
The solution of this differential equation is%
\begin{equation}
f_{m}(x)=\frac{\beta }{k_{m}^{2}}+b_{+}e^{w_{+}x}+b_{-}e^{w_{-}x},
\end{equation}%
where $w_{\pm }$ are the solutions of $[\beta
^{-1}(w^{2}-k_{m}^{2})+\varepsilon w]=0.$ The coefficients $b_{\pm }$ are
chosen such that the boundary conditions $f_{m}(0)=f_{m}(L)=0$ are
fulfilled. Extension of the solution (\ref{Sol}) to $3D$ is straightforward
from the previous calculation steps, $k_{m}^{2}\rightarrow
k_{m}^{2}+k_{n}^{2},$ where $k_{n}=n\pi /L$ introduces the discretization in
the $z-$direction.

For small forces, after performing the average over $\beta ,$ from Eq. (\ref%
{Sol}) we get%
\begin{equation}
\langle T^{(1)}\rangle (x,y)\!\simeq \!\langle \beta \rangle \tau
_{0}(x,y)+\varepsilon \langle \beta ^{2}\rangle \tau _{1}(x,y)+\varepsilon
^{2}\langle \beta ^{3}\rangle \tau _{2}(x,y),  \label{PlanoSerial}
\end{equation}%
where the auxiliary functions can be written as%
\begin{equation}
\tau _{i}(x,y)=\sum_{m=1}^{\infty }\sin (k_{m}y)g_{m}^{(i)}(x).
\end{equation}%
The addition is over odd $m,$ while%
\begin{equation}
g_{m}^{(0)}(x)=\frac{4L^{2}}{(m\pi )^{3}}\left\{ 1-\frac{\cosh \left[ \frac{%
m\pi }{2}\left( 1-2\frac{x}{L}\right) \right] }{\cosh \left[ \frac{m\pi }{2}%
\right] }\right\} ,
\end{equation}%
jointly with%
\begin{eqnarray}
g_{m}^{(1)}(x) &=&\frac{L^{2}}{(m\pi )^{3}}[\coth (m\pi )-1]e^{-\frac{m\pi x%
}{L}}\left[ (L-x)\right. \\
&&\times \left. (e^{m\pi }-e^{\frac{m\pi (L+2x)}{L}})+x(e^{2m\pi }-e^{\frac{%
2m\pi x}{L}})\right] .  \notag
\end{eqnarray}%
This function vanishes at $x=L/2,$ $g_{m}^{(1)}(L/2)=0.$ This fact follow
from the symmetries of the problem. Only for this single segment, $x=L/2,$ $%
0<y<L,$ the first correction to Eq. (\ref{PlanoSerial}) is order $%
\varepsilon ^{2}.$ The geometrical factor is%
\begin{equation}
g_{m}^{(2)}(L/2)=L^{4}\frac{\left[ 2\tanh \left( \frac{m\pi }{2}\right)
-m\pi \right] }{8m^{4}\pi ^{4}\cosh (\frac{m\pi }{2})}.
\end{equation}

\end{document}